\documentstyle[amssymb,aps,twocolumn]{revtex}
\begin{document}
\title{Low Temperature Metallic State of Ultrathin Films of Bismuth}
\author{L. M. Hernandez, Kevin A. Parendo, and A. M. Goldman}
\address{School of Physics and Astronomy, University of Minnesota, 116 Church St. SE,%
\\
Minneapolis,\\
MN 55455, USA}
\date{%TCIMACRO{\TeXButton{TeX field}{\today} }
%BeginExpansion
\today%
%EndExpansion
}
\maketitle

\begin{abstract}
Measurements of resistance vs. temperature have been carried out on a
sequence of quench-condensed ultrathin films of amorphous bismuth ({\it a-}%
Bi). The resistance below about $0.1K$ was found to be temperature
independent in a range of films with thicknesses spanning the
superconductor-to-insulator transition that would be inferred by analyzing
data obtained above $0.14K$. Film magnetoresistance was temperature
dependent in the same temperature range over which the resistance was
temperature independent. This implies that the low temperature metallic
regime is intrinsic, and not a consequence of failure to cool the electrons.
\end{abstract}


Research on ultrathin superconducting films has been focused on
superconductor-to-insulator (SI) transitions \cite{goldmarkov}, which are
believed to be quantum phase transitions (QPTs)\cite{subir}. In a QPT,
tuning an external parameter of the Hamiltonian alters the ground state. The
external parameters in previous work have been disorder, varied by changing
the thickness of quench-condensed films, and perpendicular and parallel
magnetic fields \cite{liu,hebard,yazdani,valles,Markovic,Gantmakher}.

A complication in studying QPTs is that they occur at zero temperature,
whereas data is acquired down to the lowest accessible temperatures, which
are naturally nonzero. Finite-size scaling is used to infer the existence of
the QPT, and to extract critical exponents\cite{sondhi}. A successful
finite-size scaling analysis is evidence of a QPT, but is not a proof. If
other physics intervenes at temperatures lower than those accessed in the
measurements, one may draw incomplete, if not erroneous, conclusions \cite
{Anderson}. For this reason it is essential to strive to extend measurements
to lower temperatures in search of the ``true'' ground state.

Metallic behavior at low temperatures was recently reported in studies of
the field-driven transition of homogeneous MoGe \cite{masonkapprl98} films
that would have been considered to be superconducting in the $T\rightarrow 0$
limit, based on a finite-size scaling analysis. Earlier, quench-condensed
granular films exhibited metallic behavior in zero magnetic field, beginning
at temperatures$(\thicksim 2K)$, just below those at which the resistance
initially began to drop towards zero \cite{jaeger}. In previous work on
ultrathin films of {\it a-}Bi grown on an a-Ge layer, there were no
indications of a metallic regime down to temperatures of order $0.14K$ \cite
{Markovic}.

In this letter we report measurements of $R(T,H)$ of {\it a-}Bi films over a
range of thicknesses spanning the putative SI transition. We find metallic
behavior in zero magnetic field, as evidenced by a nonzero
temperature-independent resistance at the lowest measured temperatures,
which were of order $0.050K$ \cite{temperature}. Considering only data taken
at temperatures above $0.14K$, this same set of films would appear to be
undergoing a thickness-tuned SI transition. A metallic regime was also
present when the films were subjected to magnetic fields applied in the
plane. Such fields mainly affect the amplitude of the order parameter, as
they do not introduce vortices as perpendicular fields do.

Amorphous Bi films were deposited {\it in situ} at liquid helium
temperatures onto single-crystal SrTiO$_{3}$(100) substrates, pre-coated (%
{\it in situ}) with a $6$\AA\ thick film of {\it a-}Ge \cite{Hernand}. This
was done in a chamber held at a pressure of $10^{-10}$ Torr. To prevent
annealing, substrate temperatures were held below $12K$ during growth, and
below $18K$ during other processing and handling. Film thicknesses were
increased in increments as small as $0.05$ \AA ,\ as measured using a
calibrated quartz crystal monitor. Films processed in this fashion are
believed to be homogeneously disordered \cite{Strongin}. However, a previous
experiment showed that {\it a-}Ge plays an active role in electrical
transport\cite{Nease}. Critical features of the present experiments are the
ability to change the nominal thickness of a film in tiny increments and to
grow films that are homogeneous in thickness to one part in 10$^{4}.$ This
was accomplished by using a large source-to-substrate distance (60cm) and
employing Knudsen cells as vapor sources, and maintaining ultrahigh vacuum
conditions during growth. All of the effects reported here occurred over a
nominal thickness range of order 0.2\AA\ out of approximately 9.0\AA , and
would not have been seen without such stringent control.

Resistance measurements were carried out in a Kelvinox 400 dilution
refrigerator, using either DC or AC four-probe techniques. Electrical leads
into the cryostat were filtered at room temperature using $\pi $-section
filters with a cutoff frequency of about 500 Hz. Extra filtering was
employed when a DC current source was used, bringing the cutoff frequency
down to about 10 Hz. Power dissipation in the measurement process was kept
below 1 pW. Only magnetic fields parallel to the film plane could be
applied, with values extending up to $12.5T$. In the process of insuring
adequate cooling of the films, rotation of the sample platform was
restricted to a few degrees. Parallel alignment was adjusted very carefully
at room temperature. In high fields, even a slight misalignment will result
in a substantial perpendicular field component, which could dominate the
high-field physics. For example, an error of $1^{0}$ at a field of $12T$
would result in a perpendicular component of about $0.2T$. This is greater
than the observed critical tuning field in previous studies of the SI
transition in {\it a-}Bi films\cite{Markovic}. Given the fact that one full
turn of the rotator control corresponded to about $2^{0}$, the actual error
of the alignment in this work is estimated be of order $0.1^{0}$.

The evolution of $R(T)$ of eleven films with thicknesses ranging from 8.5
\AA\ to 9.3 \AA\ is shown in Fig 1. Thinner films and thicker films, grown
in other runs (not shown) were insulating and glass-like in their responses,
or fully superconducting, respectively. Focusing on temperatures above $%
0.14K $, insulating and superconducting films appear to be clearly separated
into two groups. Although we do not show it here, resistance data of this
set of films could be collapsed in the usual manner employing a finite size
scaling form, $R\left( \delta ,T\right) =R_{c}F(\delta /T^{\frac{1}{\nu z}})$%
, with a critical exponent product $\nu z$ of $1.1\pm 0.1$ in agreement with
previous work \cite{Markovic}. The coherence length and dynamical critical
exponents are $\nu $ and $z$ respectively, and $\delta $ is the control
parameter given by $d-d_{c}$, where $d$ is film thickness and $d_{c}$ is a
critical thickness. This scaling breaks down when temperature-independent,
metallic, resistance data below $0.14K$ are included in the analysis. It
should be noted that this metallic regime was found for both insulator- and
superconductor-like films, as might be inferred from the data above $0.14K$.

In interpreting a temperature independent $R(T)$, one must consider the
possibility that as the measured temperature decreases, the electrons do not
cool. This is an important issue for disordered ultrathin films, which, as
well as being antennae for electromagnetic radiation, have tiny heat
capacities. Apart from heating by the measuring current, which is unlikely
because I-V characteristics were always linear, electromagnetic noise is
likely the major source of heating. Noise can be external or internal to the
cryostat. The latter would be Johnson noise in the electrical leads that are
in equilibrium with black-body radiation. These leads are at temperatures
higher than that of the film over much of their length. To be certain that
the electrons have cooled would require measurement of some well-understood
property of the film and its use as a thermometer, or the provision of a
separate thermometer that is in certain thermal contact with the film that
has a response to possible heating effects from the environment identical to
that of the film. Since neither of these possibilities was available, we
rely on indirect arguments. These will emerge in the course of the
discussion of $R(T,H)$, which follows.

In Fig.2, $R(H)$ at various temperatures is plotted for the $9.3\AA $\ thick
film. At the highest temperatures, in the lowest fields, $dR/dH$ is slightly
positive. With increasing field, a regime in which $dR/dH$ becomes negative
is entered. This is followed by a minimum in $R(H)$, and finally an upturn
at higher fields. In sufficiently high fields superconducting fluctuations
are completely quenched. The regime in which $dR/dH<0$ is possibly
associated with the suggestion of Kivelson and Spivak\cite{Spivak}, that a
local order parameter density can fluctuate from point to point in sign as
well as in magnitude in disordered systems near the SI transition. This can
bring about a negative magnetoresistance in a perpendicular magnetic field.
As stated previously the small error in alignment can result in a small
component of magnetic field perpendicular to the plane.

The systematics of the magnetoresistance provide evidence that the metallic
regime is intrinsic. The negative magnetoresistance grows as T decreases as
shown in Fig.2, but becomes abruptly positive when $R(T)$ becomes
temperature independent at low temperatures. It continues to increase with
further decrease of $T$. This can be seen in the inset of Fig. 2 which
emphasizes the low field regime. The magnetoresistance in the metallic
regime is a function of temperature, both in its initial change with field
and for all values of field. If the saturation of the resistance were a
consequence of electrons not cooling then these effects would not be
expected.

The upturn of $R(H)$ in high parallel fields is hard to explain
quantitatively as it is a consequence of the combined action of the large
parallel component of the field and the very much smaller, but unknown
perpendicular component. A linear dependence of $R(H)$ on field would be
expected if there were flux flow resistance due to a perpendicular field
component \cite{Markovic1}. It should be noted that at high fields there is
a significant contribution to the resistance which is quadratic with field.
This could be associated with the weakening of the order parameter amplitude
fluctuations by the parallel component of field.

Further interesting effects are seen if the temperature dependence of the
resistance is plotted for fields above the magnetoresistance minimum. The
quenching of superconducting fluctuations in this regime is shown in Fig. 3
for the $9.19\AA $ thick film. Its low temperature behavior is always
metallic. As the field is increased, the domain in temperature over which $%
dR/dT=0$ broadens significantly and reaches a maximum width of about $100mK$%
.. This maximum width can be seen much more dramatically in Fig. 4 which
shows data from the $9.09\AA $\ thick film, in which the temperature of the
onset of the metallic regime is plotted as a function of its resistance. For
this film, the resistance of the widest metallic regime was $12,900\Omega $,
which is remarkably close to twice the value of the quantum resistance for
pairs. This resistance is actually equal to the zero-field resistance of the
film in the low temperature limit. This particular film is the first of
those in the sequence of films shown in Fig. 1 that exhibits superconducting
fluctuations as evidenced by a downturn in $R(H)$ at higher temperatures.
All of the films exhibiting superconducting fluctuations display a similar
response to an applied parallel magnetic field. It is important to note that
the systematics of this variation of resistance with field are different
from what was reported by Mason and Kapitulnik\cite{MasonandKapitulnik}{\bf %
.. } This may be a consequence of the fact that the field in this work was
parallel rather than perpendicular to the film plane.

One may compare the present results with those obtained in studies of
quench-condensed {\it a-} Ga films which behaved in a manner very similar to
the films of the present work, but became metallic at temperatures as high
as $2K$ \cite{jaeger}. In that work, the temperatures at which $dR/dT$ first
fell to zero were dependent on the measuring current. In the subsequent work
of Ref.\cite{Christiansen}, on similar films, it was proposed that the
nonlinearities and metallic behaviors in films that appeared to become
insulating at high temperatures were a consequence of charge motion, whereas
for those that appeared to become superconducting, they were due to vortex
motion. This was arrived at because the systematics of the data suggested
that insulating and superconducting states at $T=0$ would only be found in
the limit of zero measuring current. Although the I-V characteristics were
nonlinear, the systematics of the data could also be used to eliminate the
possibility of heating. It is conceivable that similar nonlinear features of
the I-V characteristics for {\it a-}Bi films exist, but their observation
would require much lower current levels than were employed in the
measurements.

Metallic regimes at low temperatures in ultrathin films that exhibit local
superconductivity have been the subject of several theoretical works, which
emphasize the bosonic nature of these systems as well as, in some instances,
the role of dissipation \cite{Doniach,Lee,Phillips,Kapitulnik}. The data we
have presented cannot validate any of these theories, some of which make
very specific predictions, such as glass-like behavior\cite{Phillips}. On
the other hand, the data can be used to make a case for the metallic regime
being intrinsic and not a consequence of failure to cool the electrons.

The values of either thickness or magnetic field that result in maximally
wide metallic regimes could be evidence of the metallic quantum critical
point of the SI transitions proposed in the context of phase-only Bose
Hubbard models \cite{Fisher,Wallin}. We conjecture that when the tuning
parameter of the quantum phase transition is close to its critical value,
evidence of quantum critical fluctuations, in this case, a metallic regime,
extends to high temperatures, resulting in behavior such as that exhibited
in Fig. 4. In this scenario metallic behavior at other fields (or
thicknesses) is seen at nonzero temperatures because the free motion of
vortices and charges that respond to the measuring current masks the
superconducting and insulating ground states, as discussed above. The actual
value of the resistance of the separatrix for the thinnest film exhibiting
superconducting fluctuations is very close to $h/2e^{2}$ which is double the
quantum resistance for electron pairs. It turns out that this is the value
of the critical resistance obtained from finite-size scaling of Monte Carlo
data in simulations of the phase-only Bose Hubbard model\cite{Wallin}. This
scenario implies that there is a quantum critical point separating
insulating and superconducting phases, with the metallic regime that
obscures it resulting from a nonzero measuring current. Alternatively, there
might not be a quantum critical point at all. In this instance, the striking
magnetic field dependence of the temperature onset of the metallic regime
such as shown in Fig. 4 would be just a feature of the phase diagram of the
putative Bose metal regime.

In summary, the extension of measurements of $R(T)$ of ultrathin,
quench-condensed films of {\it a-}Bi down to temperatures of order $0.05K$
have revealed metallic behavior over a range of film thicknesses near the SI
transition. This suggests that the apparent superconducting and insulating
ground states inferred from the analysis of data obtained at temperatures
above $0.14K$ and discussed in many theories, may not be realized. Evidence
was presented supporting the results being intrinsic, and not a consequence
of the electrons in the film not being cooled. Lower temperatures and lower
measuring current densities may be needed to find the ``true'' ground state,
at least for films in the transition regime between insulating and
superconducting behavior.

This work was supported by the National Science Foundation Condensed Matter
Physics Program under grant Grant DMR-0138209.

% now the references. delete or change fake bibitem. delete next three
%   lines and directly read in your .bbl file if you use bibtex.
....

\begin{figure}[tbp]
\caption{Evolution of $R(H)$ at various temperatures for the 9.3\AA\ thick
film, with H in-plane. Temperatures, for curves from top to bottom are: 500,
250, 200, 150, 100, and 50mK. Inset: the same data plotted as percentage
change in the resistance vs. field. The range is limited to 1 T, so as to
highlight the low-field behavior. The two upper curves were obtained at 100
mK and 50 mK, which are in the metallic regime.}
\label{fig2}
\end{figure}
\begin{figure}[tbp]
\caption[Evolution of$R(T)$ of the 9.19\AA\ film as a function of in-plane
magnetic field. Field values from top to bottom are: 12.5, 12, 11.6, 11.5,
11, 10, 9, 8, 7, 6, 5, 4, 3, 2, and 0 T. Inset: temperature at which dR/dT
becomes zero is plotted vs. applied field.]{Evolution of $R(T)$ of the
9.19\AA\ film as a function of in-plane magnetic field. Field values from
top to bottom are: 12.5, 12, 11.6, 11.5, 11, 10, 9, 8, 7, 6, 5, 4, 3, 2, and
0 T. }
\label{fig3}
\end{figure}
\begin{figure}[tbp]
\caption[Evolution of$R(T)$ of the 9.19\AA\ film as a function of in-plane
magnetic field. Field values from top to bottom are: 12.5, 12, 11.6, 11.5,
11, 10, 9, 8, 7, 6, 5, 4, 3, 2, and 0 T. Inset: temperature at which dR/dT
becomes zero is plotted vs. applied field.]{Plot of temperature T$_{flat}$
at which $R(T)$ becomes temperature independent vs. the value of that
resistance for the $9.09\AA $ thick film. A sharp peak at a sheet resistance
of 12,900 $\Omega $ is evident. }
\label{fig4}
\end{figure}
%figures follow here
%Fill in the caption in the braces of the \caption{} command. Put the label
%\begin{figure}
% \caption{}
% \label{}
% \end{figure}
%Here is an example of the general form of a table:
%that you will use with \ref{} command in the braces of the \label{} command.
%Insert the column specifiers (l, r, c, d, etc.) in the empty braces of the
%\begin{tabular}{} command.
% \begin{table}
% \caption{}
% \label{}
%\begin{tabular}{}
% \end{tabular}
% \end{table}


\begin{references}
\bibitem{temperature}  The actual low temperature limit on the sample holder
was close to 0.020 K when the refrigerator was cooled to its base
temperature. This was determined from other work using a calibrated
thermometer mounted on it. The calibration for the thermometer used for the
present studies had a lower limit of 0.050K, hence the use of the term
``measured temperature.'' All of the data shown refers to the latter, but
the metallic regime discussed here actually persisted to the temperature
reached when the refrigerator was allowed to cool to its base temperature.
\end{references}
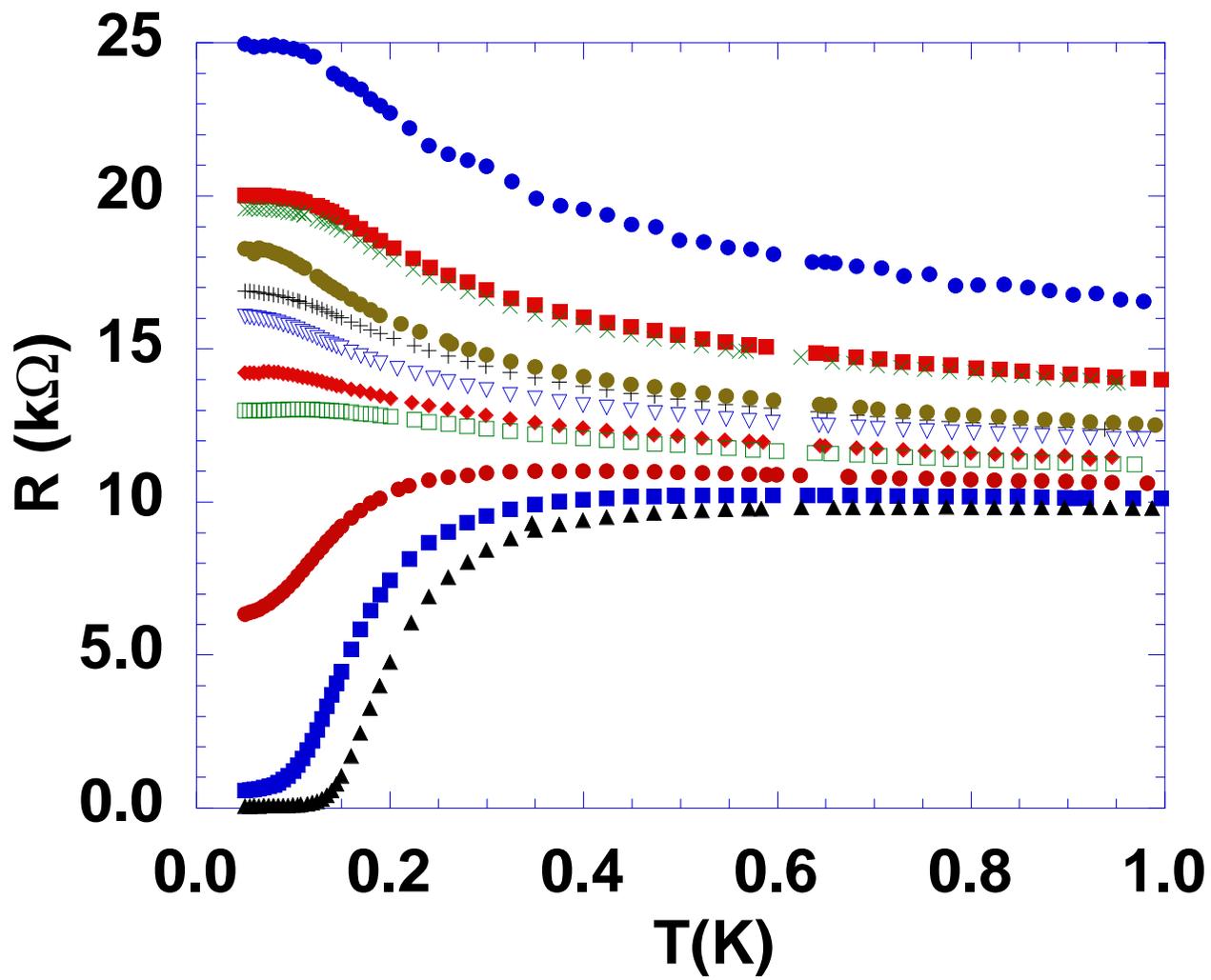
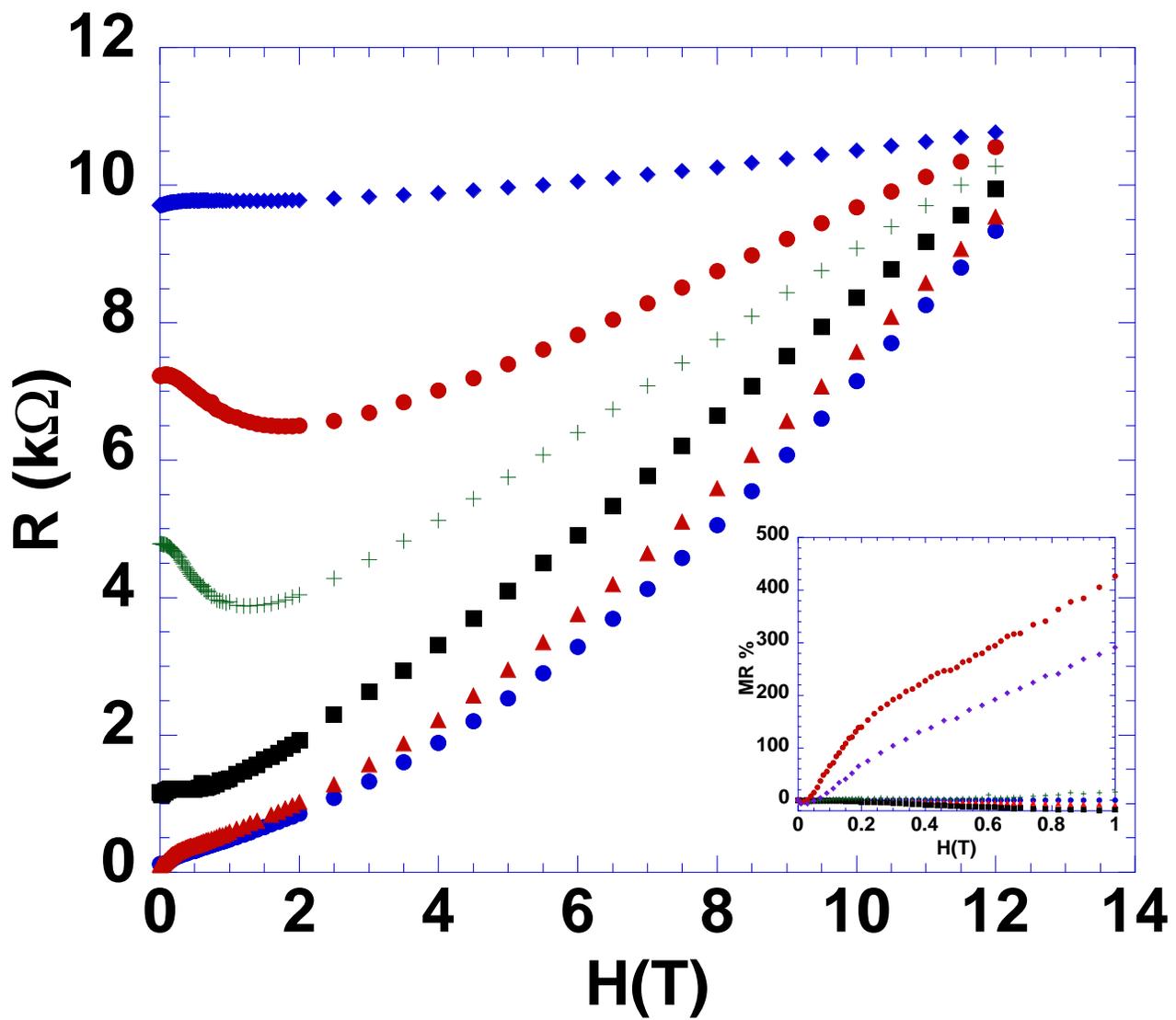
\end{document}